# Title: Full tomography of topological Andreev bands in graphene Josephson junctions

**Authors:** Woochan Jung[1]†, Seyoung Jin[1]†, Sein Park[1], Seung-Hyun Shin[1], Kenji Watanabe[2], Takashi Taniguchi[3], Gil Young Cho[1]*, Gil-Ho Lee[1,4]*

**Affiliations:**

[1]Physics Department, POSTECH, Pohang 37673, Republic of Korea.

[2]Research Center for Functional Materials, National Institute for Materials Science, Tsukuba, Japan.

[3]International Center for Materials Nanoarchitectonics, National Institute for Materials Science, Tsukuba, Japan.

[4]Asia Pacific Center for Theoretical Physics, Pohang, Republic of Korea.

† These authors contributed equally to this work

*Corresponding author. Email: gilyoungcho@postech.ac.kr or lghman@postech.ac.kr

**Abstract:** Multiply connected electronic networks threaded by flux tubes have been proposed as a platform for adiabatic quantum transport and topological states. Multi-terminal Josephson junction (MTJJ) has been suggested as a pathway to realize this concept. Yet, the manifestations of topology in MTJJ remain open for experimental study. Here, we investigated the artificial topological band structure of three-terminal graphene Josephson junctions. Employing tunnelling spectroscopy and magnetic flux gates, we captured the tomography of the Andreev bound state (ABS) energy spectrum as a function of two independent phase differences. This ABS spectrum exhibits the topological transition from gapped to gapless states, akin to the band structure of nodal-line semimetals. Our results show the potential of graphene-based MTJJs for engineering band topologies in higher dimensions.

## Introduction

The Josephson junction (JJ) stands as a pivotal component in quantum technology and topological condensed matter physics. Proximity JJs, comprising superconductor/normal-metal/superconductor heterostructures, provide a versatile platform for harnessing the distinct properties of normal metals alongside the coherent superconducting order. Within the junction, electronic bound states form, known as Andreev bound states (ABSs), which govern its behavior. Thus, the controllability of ABSs is paramount for its quantum technology applications, including superconducting qubits (*1-7*), quantum sensors (*8-10*), parametric amplifiers (*11, 12*). The energy spectrum of an ABS sensitively depends on the superconducting phase difference between the superconductors, evolving smoothly with changes in this phase difference. This spectral evolution bears resemblance to the "band structures" of higher-dimensional crystalline systems. By substituting phase differences with the quasi-momentum of higher-dimensional lattice systems, we can identify the ABS spectrum as a band structure of higher spatial dimensions. Consequently, they are termed

"Andreev band structures" in JJs. This distinctive mapping presents a novel and controllable avenue for realizing higher-dimensional topological band structures, which may not be immediately accessible in usual condensed matter systems, within mesoscopic-sized JJ devices.

The accessible band topology and its characteristics in an actual device heavily rely on the detailed nature of the junction, including factors such as the number of superconducting terminals, conducting channels, symmetry of the normal region, and available probes. Conventional two-terminal JJs pose particular limitations; ABSs in the JJ fail to intersect zero energy unless a unique phase difference ($\varphi = \pi$) is achieved, and even then, only a system with finely tuned perfect transmission can exhibit this phenomenon. Therefore, the finely tuned nodal crossing, even if it occurs in the two-terminal JJ, is not associated with topological invariants (*13-16*). In contrast, multi-terminal Josephson junctions (MTJJs) offer an Andreev band with robust topological features (*17-24*). Not only do they feature band crossings, but they can also be associated with topological invariants, such as the presence of non-contractible loops or the winding of the superconducting order parameter (*25-30*). Despite the growing theoretical interest in MTJJs, previous studies (*31-41*) have typically focused on transport features rather than directly proving the ABSs, thus limiting the exploration of exotic band topology. While recent experimental works have aimed to directly prove the Andreev band of MTJJs (*42-47*), stringent requirements, including phase coherence among all terminals and high-energy-resolution spectroscopy with full control of phase differences, have not been fully met.

To address these limitations, we developed a highly tunable three-terminal graphene JJ [Fig. 1**A**] and conducted a detailed investigation of its two-dimensional Andreev band structure using a superconducting tunnel probe. Graphene serves as the normal-scattering region, enabling phase-coherent transport. In addition, our device permits independent control of each superconducting phase, allowing us to explore the Andreev band at any point in the two-dimensional "phase space" [Figs. 1**B**, **C**]. By combining this comprehensive tomography of the electronic structure with theoretical model calculations, we demonstrate that the junction aligns with the predictions of random matrix theory and features nodal lines [Fig. 1B] atop the topological parity transitions (*28*).

**Results**

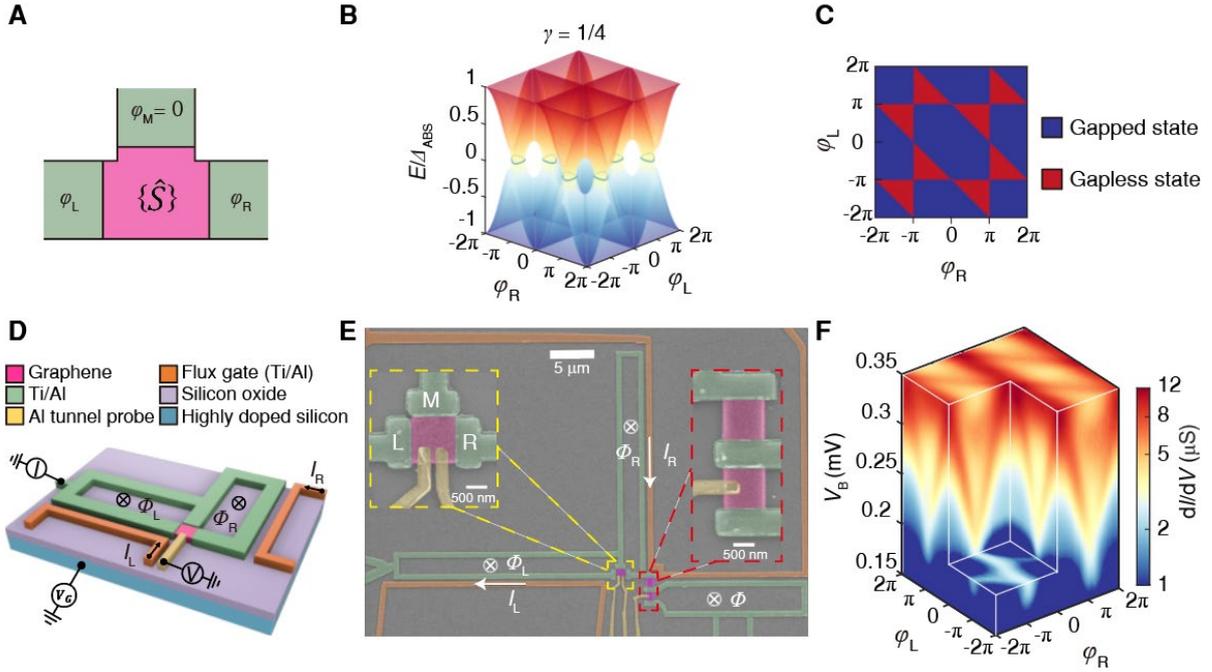

**Fig.1. Andreev band in three-terminal JJ (3TJJ) and device schematic.** (**A**) 3TJJ across normal-scattering region (magenta), which is described by scattering matrix $\{\hat{S}\}$. Each terminal has superconducting phase $\varphi_i$ ($i =$ M, L, R). (**B**) Andreev band of isotropic 3TJJ for $\gamma = 1/4$ spanned by two superconducting phase difference variables that can be tuned by external magnetic flux, $\Phi_L$ and $\Phi_R$. Here, $\gamma$ is value that parameterizes transmission probability in scattering matrix. Green lines represent nodal lines formed by band crossing at zero energy. (**C**) Phase diagram of density of states (DOS) at zero energy. (**D**) Schematics of gate-tunable 3TJJ integrated with two superconducting loops. External magnetic flux, $\Phi_L$ and $\Phi_R$, are induced by bias current flowing through flux lines, $I_L$ and $I_R$, respectively. (**E**) False-colored scanning electron microscope image of device. Left inset, graphene region (magenta) is in ohmic contact with three aluminum (Al) superconducting terminals (green) that form superconducting loops and are in tunnel contact with other Al electrodes (yellow). Right inset, magnified image of neighboring graphene JJs. (**F**) Experimentally measured differential conductance $dI/dV$ plotted as function of bias voltage $V_B$ of superconducting tunnel probe, $\varphi_L$ and $\varphi_R$, at fixed gate voltage $V_G = 20$ V.

The three-terminal Josephson junction (3TJJ) comprises three independent superconducting terminals, each characterized by its superconducting phase $\varphi_i$ (i = M, L, R), and a central normal-scattering region, as shown in Fig. 1A. These superconductors are coupled through the graphene region, facilitating the flow of Josephson supercurrents between them. The Josephson supercurrent within the 3TJJ is determined by the Andreev band, whose energy spectrum depends on the phase differences of the superconducting electrodes (*22*). Due to gauge invariance, we can set one of the superconducting phases, $\varphi_M$, to zero without loss of generality. Then, two phase differences $\varphi_{R/L} = 2\pi\Phi_{R/L}/\Phi_0$ are physically independent, which can be modulated by the external magnetic fluxes $\Phi_{R/L}$ [Figs. 1D, E]. Here, $\Phi_0$ is the superconducting flux quantum. The independent control of $\varphi_{R/L}$ allows us to tune the two-dimensional Andreev bands, the detailed structure of which is the focus of this study.

The Andreev band of the 3TJJ is well-described by the scattering matrix formalism, involving a scattering matrix $\{\hat{S}\}$ that relates the incoming and outgoing wave functions in the junction (*17, 28, 48*). Although several additional details need to be addressed to make a sensible comparison between theory and experiment, we can already discern the essential features of the Andreev bands from the simplest model, namely an isotropic, two-channel ballistic 3TJJ in the short-junction limit [see Supplementary, S1-1B]. The model's scattering matrix has a single free parameter $\gamma \in [0,1]$, parametrizing the transmission probability, which determines the entire band structure. Figure 1**B** depicts a typical Andreev band of the model with $\gamma = 1/4$, plotted as a function of the two-phase differences, $\varphi_L$ and $\varphi_R$, which can be adjusted by varying the external magnetic fluxes $\Phi_L$ and $\Phi_R$. Notably, the Andreev band exhibits crossings at zero energy, giving rise to the formation of nodal lines (green lines). Such a feature holds for $\gamma \in [0, 1/3)$. The low-energy spectrum near the nodal lines reveals a linear dispersion [see Supplementary, S1-1A], an important characteristic of topological nodal-line semimetals, extensively studied in crystalline material systems (*49-51*). Although it is not our immediate focus, we note that the same 3TJJ device structure under appropriate Floquet driving was theoretically suggested to simulate the topological physics of the Haldane model on a honeycomb lattice (*52*). To be more realistic than the above toy model, we must consider many conducting channels with different details in the scattering region, the number of which is roughly $2W/\lambda_F \sim 80$. $W \sim 1$ μm is the contacted width of graphene and $\lambda_F \sim 25$ nm is the Fermi wavelength in graphene [see Supplementary, S6]. This implies that the observed tunneling conductance in the experiment should be regarded as the sum over a set of many different scattering matrices $\{\hat{S}\}$. Such a junction is described by random matrix theory (*17, 21, 53-56*), which governs the allowed regions in the space of $(\varphi_L, \varphi_R)$ for the Andreev band to touch the zero energy [see Supplementary, S1-2]. Consequently, this results in the "phase diagram" of gapped and gapless states in $(\varphi_L, \varphi_R)$ as shown in Fig.1**C**, consistently captured by our toy model. As $\gamma$ is continuously tuned from 0 to 1/3, the line nodes completely fill inside the allowed zero energy region [red regions in Fig.1**C**].

We constructed graphene-based 3TJJ devices embedded with two superconducting loops and probed the Andreev band using a superconducting tunnel probe [Figs. 1**D, E**]. Three Aluminum (Al) superconducting terminals (green color) were contacted to the edge of the graphene with a titanium (Ti) adhesion layer. Two superconducting rings were formed to independently control the phase differences $\varphi_{R/L}$ by injecting a current $I_{R/L}$ into the flux gates (orange color) and producing a magnetic flux $\Phi_{R/L}$ threading each ring. The graphene was encapsulated with top and bottom hexagonal boron nitride insulating layers to minimize impurity scatterings within the graphene and maintain it as a ballistic conductor [see Materials and methods]. We achieved the short-junction limit by designing the ratio of channel length ($L$) to superconducting coherence length ($\xi$) to be $L/\xi \sim 0.26$, smaller than 1. A highly doped silicon substrate served as a global backgate for controlling the carrier density of the graphene. To form a superconducting tunnel contact, Al superconducting electrodes [yellow color in Figs. 1**D, E**] were directly contacted to the graphene edge without any adhesion layers, resulting in a tunneling barrier between Al and graphene due to a large interatomic distance (*57*). Figure 1**F** shows the differential tunneling conductance ($dI/dV$) map as a function of $(\varphi_L, \varphi_R)$ and bias voltage $V_B$. All measurements were conducted at the base temperature of 20 mK unless stated otherwise. JJs with and without a superconducting loop [Fig.1**E** (right inset)] were measured to confirm the short ballistic nature of the junction [see Figs. S3 and S4 for details].

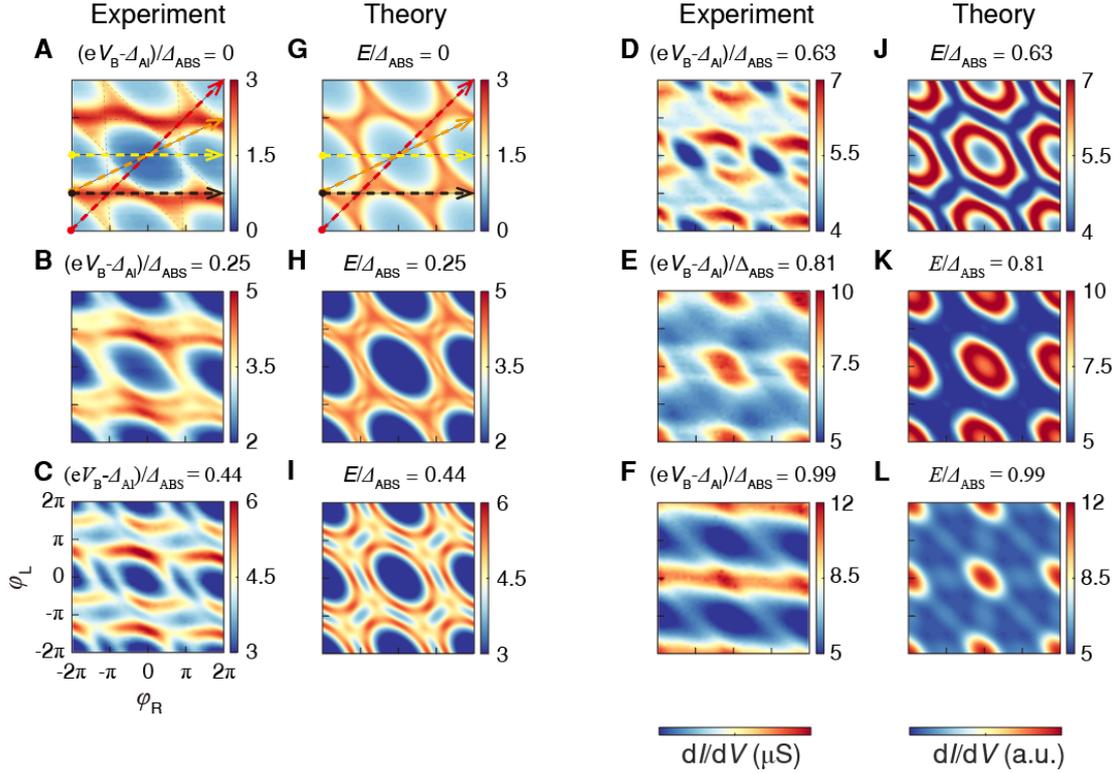

**Fig. 2. Tomography of Andreev band.** (**A-F**) Experimentally measured differential tunneling conductance, $dI/dV$, as a function of two-phase differences $(\varphi_L, \varphi_R)$, for different normalized energies $(eV_B - \Delta_{Al})/\Delta_{ABS}$, at gate voltage $V_G = 20$ V. (**G-L**) Theoretically calculated $dI/dV$ as function of $(\varphi_L, \varphi_R)$ for different normalized energy $E/\Delta_{ABS}$. In **A** and **G**, the color of dashed arrows corresponds to line cuts that are displayed in Fig. 3.

Figs. 2**A-F** depict iso-energy cuts of the $dI/dV$, approximately representing the DOS of the Andreev band in the 3TJJ, at different normalized energies $(eV_B - \Delta_{Al})/\Delta_{ABS}$. The energy is normalized by the induced superconducting gap ($\Delta_{ABS} = 0.155$ meV) after subtracting the offset due to the superconducting gap of the Al tunnel probe ($\Delta_{Al} = 0.19$ meV) [see Fig. S5 for further details]. This representation is analogous to angle-resolved photoemission spectroscopy (ARPES) measurements, which visualize the band structure as a function of quasi-momentums.

In the theoretical calculations, we employed the scattering matrix formalism with random matrix theory to describe our 3TJJ in the short-junction limit. The spin-rotational symmetry and time-reversal symmetry in the normal graphene region necessitated sampling the scattering matrices $\{\hat{S}\}$ from the circular orthogonal ensemble. We sampled only the random scattering matrices that describe highly transparent junctions [see Supplementary, S1, 2]. The calculated $dI/dV$ is presented in Figs. 2**G-L**, representing the result of averaging over 50,000 samples of the scattering matrices. There is no anisotropic bias in sampling scattering matrices [see Supplementary, S2] between the three superconducting terminals, corresponding to a situation where all three terminals in the 3TJJ are equally coupled (on average) to each other.

Figs. 2**A** and 2**G** illustrate the evolution of the (experimental) $dI/dV$ and (theoretical) $dI/dV$ at zero energy, respectively. In this two-dimensional phase space, transitions between the gapless (red region) and gapped states (blue region) can be distinguished. Notably, these gapped regions are separated by gapless regions, forming closed loops. This feature is consistent with the phase diagram obtained from our model calculations, as shown in Fig. 1**C**.

As $E/\Delta_{ABS}$ increases, the gapless state (red region) at zero energy splits into pairs, and simultaneously, the closed loops begin to contract over the phase space. These loops converge to points as $E/\Delta_{ABS}$ approaches 1. When compared to the theory [Figs. 2**G-L**], there is a small anisotropy in experimental data [Figs. 2**A-F**], showing enhanced $dI/dV$ contrast along the $\varphi_L$ direction compared to that along the $\varphi_R$ direction. This is attributed to the anisotropic contact transparency inherent in our device. In particular, the connectivity of terminal R was slightly lower than that of terminals L and M. This leads to a Josephson coupling between terminals M and L, which is slightly stronger than that between terminals M(L) and R [see Fig. S8 and Supplementary, S2-1 for details]. Except for this minor discrepancy, our theoretical calculations aptly captured the main features evident in the measurements displayed in Figs. 2**A-F**. Therefore, we can conclude that the hybridization of the three highly transmissive ABSs exists in the graphene region of the 3TJJ. Furthermore, the gapless regions represent the manifestation of states crossing at zero energy, a phenomenon consistently observed at other gate voltages [see Fig. S6]. Interestingly, each gapped region can be associated with a topological index, namely the superconducting vorticities in the left and right superconducting loops (*28, 43*). Hence, by modulating the external fluxes $(\Phi_L/\Phi_0, \Phi_R/\Phi_0)$, the topological transition between these gapless and gapped regimes in equilibrium can be induced, which is well demonstrated in our experimental data [Figs. 2**A-F**].

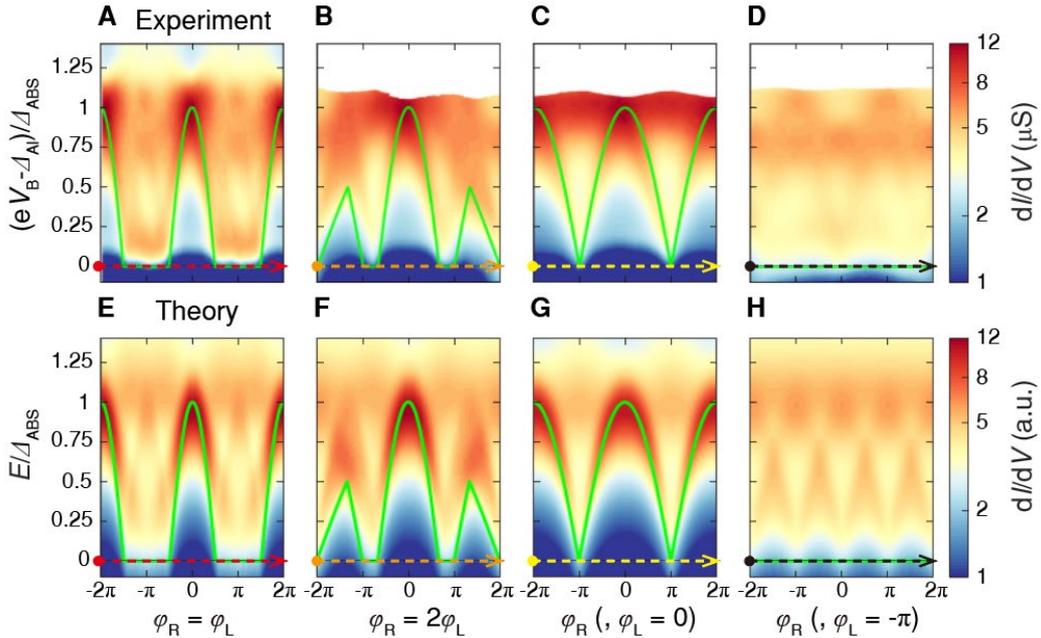

**Fig. 3. Energy spectrum of Andreev band.** Experimentally measured differential tunneling conductance, $dI/dV$ (upper panels, **A-D**), and theoretically calculated $dI/dV$ (lower panels, **E-H**) are presented as functions of normalized energy, $(eV_B - \Delta_{Al})/\Delta_{ABS}$ and $E/\Delta_{ABS}$. These measurements were taken along phase-space line cuts (indicated by the horizontal dashed arrow lines) in plots shown in Fig. 2**A** and 2**G**, respectively. Green solid lines represent analytical lower bound of Andreev band, which refers to theoretical prediction for minimum energy gap within short junction. In panel **A**, $dI/dV$ versus $(eV_B - \Delta_{Al})/\Delta_{ABS}$ along red dashed line defined by $\varphi (= \varphi_L = \varphi_R)$ was measured with external superconducting solenoid magnet.

To observe more details of the topological transition from a gapped region to a gapless region, the bias-voltage dependence of the spectrum along various directions in the phase space is plotted in Fig. 3. The energy spectrum not only enables the examination of the dispersion relation in the phase space but also facilitates the study of the properties of the complete Andreev spectrum. To demonstrate this, we selected specific line cuts in the phase space [indicated by the colored dashed arrows in Fig. 2**A** and 2**G**] and plotted the experimental $dI/dV$ [Figs. 3**A-D**] and theoretically calculated $dI/dV$ [Figs. 3**E-H**] along these cuts. The green solid lines depicted in Fig. 3 represent the analytical lower bound of the Andreev band in the short-junction limit, indicating that there is no state with energy below the green solid line [see Supplementary, S1-2]. Although the overall trend matches between experimental and theoretical trends, a nonzero differential tunneling conductance below the lower bound is observed [see Figs. 3**A-C**]. This is attributed to the finite energy resolution of our tunneling probe, estimated to be approximately 0.02 meV [see Fig. S9], and also to the deviation from the strict short-junction limit (*28*). Nonetheless, the theoretical calculations and experimental data agreed well, as we discuss in detail below.

First, the locations of the gapless regions agree well between the theory and experiment. Whenever zero energy crossings are allowed by the lower bound, the experimental $dI/dV$ develops at zero energy, mirroring the theoretical simulation results. For example, Fig. 3**A** represents the sweep in the direction of the red dashed arrow in Fig. 2**A**. This direction, defined by $\varphi$ $(= \varphi_\text{L} = \varphi_\text{R})$, can be achieved even without the flux gates and has been explored previously (*43*). Notably, the two distinct gapped and gapless regimes become evident in the vicinity of zero energy, which appears for $-1.5\pi \leq \varphi \leq -0.5\pi$ and $0.5\pi \leq \varphi \leq 1.5\pi$. For $1.5\pi \leq |\varphi| \leq 2\pi$ and $|\varphi| \leq 0.5\pi$, the Andreev band is in the gapped region as clearly observable both in theory and experiment [Figs. 3**A**, **E**]. The agreement between the theoretical and experimental data on the locations of the gapless regions and the intensities of DOS is also apparent in other line cuts [Figs. 3**B-D**]. Finally, at the low-energy window (below approximately 0.25 meV), the energy-phase dispersion matches with the analytical lower bound, as depicted by the green solid line [see Figs. 3**A-C**]. These green solid lines agree with the linear dispersion relation of the outermost nodal line of the Andreev band [see Supplementary, S1-2A], which was reasonably well resolved in our experiment.

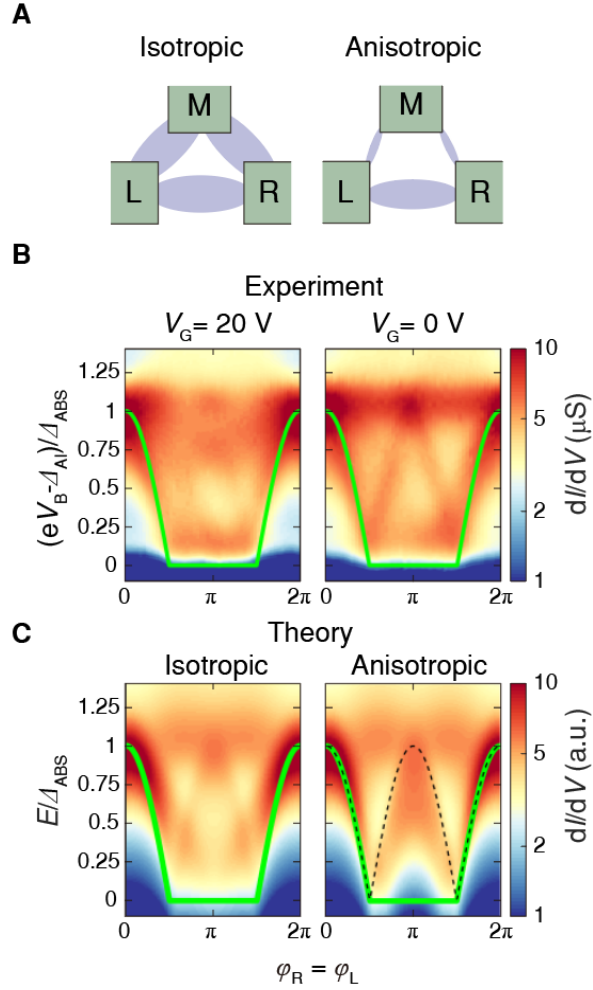

**Fig. 4. Gate voltage, $V_G$, dependence on anisotropy of terminal connectivity.** (**A**) Schematics of the device model for elucidating the anisotropic nature. Three superconducting terminals (green) are interconnected through three Josephson couplings (blue oval), strength of which is adjusted with $V_G$. Left panel depicts situation when all terminals have isotropic connectivity, while right panel depicts relatively weak connectivity of terminal M. (**B**) Experimental measurements of tunneling differential conductance, $dI/dV$ at $V_G = 20$ V (left panel) and $V_G = 0$ V (right panel) plotted against $V_B$ along phase-space linecuts, as indicated by red dashed arrows in Fig. 2**A** and 2**G**. Green solid lines represent analytical lower bound of Andreev band, which refers to theoretical prediction for minimum energy gap of short junction. (**C**) Theoretical calculations of $dI/dV$ for both isotropic (left panel) and anisotropic (right panel) cases. Black dashed line exposes dehybridization feature transitioning from three-terminal Andreev band to two-terminal Andreev band when connectivity of terminal M is weakened.

In the same device, we could control the anisotropy in the Josephson coupling strength among the three terminals by controlling the Fermi level of graphene via in-situ electrostatic gating [Fig. 4**A**]. Figure 4**B** presents $dI/dV$ measured at two distinct gate voltages, $V_G = 20$ V and $0$ V, along the symmetric line in the phase map, $\varphi\ (=\varphi_L = \varphi_R)$. For both gate voltages, we observed a clear oscillation of the $dI/dV$ peak with $\Phi_0$ periodicity, which aligns well with the theoretical prediction depicted by the green solid line. At a gate voltage of $0$ V, near the charge neutrality point ($V_{\text{Dirac}} = -3$ V), a distinct feature of the lambda ($\Lambda$)-shape enhancement of $dI/dV$ centered around $\varphi = \pi$ emerges with a periodicity of $\Phi_0/2$.

This gate dependence can be understood intuitively via a simplified model shown in Fig. 4**A**. In theory, the strength of the Josephson couplings can be modulated by adjusting the transmission probabilities in the scattering matrix [see Supplementary, S2-1]. The calculated result for the isotropic case presented in the left panel of Fig. 4**C** is well matched to the experimental observations in the left panel of Fig. 4**B**. For an anisotropic case where the terminal M has relatively weak connectivity compared to the others [right panel of Fig. 4**A**], the Josephson coupling between terminals L and R with the phase difference $\varphi_L + \varphi_R = 2\varphi = 2\pi(2\Phi)/\Phi_0$ dominates, resulting in $\Phi_0/2$-periodicity of the Andreev band. Thus, the calculation shows more pronounced $dI/dV$ along the black dashed line as shown in the right panel in Fig. 4**C**, which is well captured by the lambda (Λ)-shape enhancement in $dI/dV$ around $\varphi = \pi$ in the right panel of Fig. 4**B**. Another distinctive feature is the reduction of the DOS that spreads from low energy to high energy around $\varphi = \pi$. These features signify a transition from a three-terminal Andreev band to a two-terminal band as the connectivity of terminal M weakens [see Fig. S7]. The good agreement between the theory and experiment evidences the control of Andreev bands in the 3TJJ by utilizing the backgate with geometric anisotropy, where the three-terminal Andreev band is dehybridized into a two-terminal Andreev band.

**Conclusions**

Tunneling spectroscopy revealed that the Andreev band was localized within the graphene region of the 3TJJ as a function of two independent phases. By integrating extensive tomography of the Andreev band structure with theoretical model calculations, we demonstrated that the Andreev band is consistent with the predictions of random scattering matrix theory, notably exhibiting nodal lines superimposed on topological parity transitions. Additionally, a key aspect of our research is the *in-situ* controllability of multi-terminal couplings through electrostatic gating, which enabled us to observe dehybridization from a three-terminal Andreev band to a two-terminal band. Our study elucidates the Andreev band structure of the 3TJJ and opens new avenues for engineering the band topology.

In the near future, when introducing spin-orbit coupling, 3TJJ may emerge as promising platforms for a new type of superconducting spin qubits and for generating Majorana bound states (*28*). While the nodal-line features of Andreev band were discussed in this study, other topological aspects of the 3TJJ could be investigated further either by introducing a magnetic flux through the junction area (*23*) or by applying elliptically polarized electromagnetic waves to generate Floquet states (*52, 58*). Moreover, incorporating additional superconducting terminals (*17*) or tuning the anisotropy of terminal connectivity (*29*) could enhance our capabilities in Andreev band engineering. These pathways not only facilitate the fundamental study of topology in higher dimensions even beyond three but also pave the way for Josephson junction-based quantum technology applications.

**Materials and Methods**

**Device fabrication**

Graphene and hexagonal boron nitride (hBN) flakes were exfoliated on separate silicon oxide wafers. To encapsulate a graphene monolayer, a dry transfer method was employed with hBN flakes. The top hBN was 56 nm thick, and the bottom hBN was 55 nm thick. The graphene stack was shaped using reactive ion etching with $CF_4$ and $O_2$ plasmas, while the electrodes were patterned using electron-beam lithography and deposited via electron-beam evaporation onto the freshly etched edges of graphene. Ohmic-contact electrodes comprised a 5-nm titanium adhesion layer deposited at an angle of ±30° to ensure good contact, followed by a 100-nm Al superconducting layer. The Al layer was deposited using BN-TiB2 crucibles at a rate of 1 Ås$^{-1}$. Additionally, a flux gate was deposited simultaneously with the ohmic-contact electrodes. A superconducting tunnel probe electrode, comprising an 80-nm Al layer without any adhesion layers, was deposited using BN-TiB2 crucibles at a rate of 0.3 Ås$^{-1}$. The evaporation chamber pressure during the evaporation was maintained at less than $1 \times 10^{-7}$ mTorr.

**Acknowledgments:** We thank E. G. Arnault for reading the manuscript. **Funding:** This work was supported by National Research Foundation (NRF) grants (Nos. 2021R1A6A1A10042944, 2022M3H4A1A04074153, RS-2023-00207732, RS-2023-00208291, 2023M3K5A1094810, and 2023M3K5A1094813) and ITRC program (IITP-2022-RS-2022-00164799) funded by the Ministry of Science and ICT. Additionally, support was received from the Air Force Office of Scientific Research under award number FA2386-22-1-4061, Institute of Basic Science under project code IBS-R014-D1, Samsung Science and Technology Foundation under projects (Nos. SSTF-BA1702-05, SSTF-BA2101-06 and SSTF-BA2002-05), and Samsung Electronics Co., Ltd. (IO201207-07801-01). K.W. and T.T. acknowledge support from the JSPS KAKENHI (Grant Numbers 21H05233 and 23H02052) and World Premier International Research Center Initiative (WPI), MEXT, Japan. **Author contributions:** G.-H.L. and G.Y.C. conceived and supervised the project. W.J. designed and fabricated the devices. T.T. and K.W. provided the hBN crystal. W.J, S.P. and S.-H.S. performed the measurements. S.J. and G.Y.C. carried out theoretical calculations. W.J., S.J., G.Y.C. and G.-H.L. performed the data analysis and wrote the paper. **Competing interests:** None declared. **Data and materials accessibility:** All additional data required to assess the conclusions of the paper are available in the paper itself or in the supplementary materials.

# Supplementary Materials for

## Full tomography of topological Andreev bands in graphene Josephson junctions

Woochan Jung[1]†, Seyoung Jin[1]†, Sein Park[1], Seung-Hyun Shin[1], Kenji Watanabe[2], Takashi Taniguchi[3], Gil Young Cho[1]*, Gil-Ho Lee[1,4]*

[1]Physics Department, POSTECH, Pohang 37673, Republic of Korea.

[2]Research Center for Functional Materials, National Institute for Materials Science, Tsukuba, Japan.

[3]International Center for Materials Nanoarchitectonics, National Institute for Materials Science, Tsukuba, Japan.

[4]Asia Pacific Center for Theoretical Physics, Pohang, Republic of Korea.

† These authors contributed equally to this work

*Corresponding author: Email: gilyoungcho@postech.ac.kr or lghman@postech.ac.kr

**The PDF file includes:**

>Supplementary Text
>Figs. S1 to S10
>References

# Table of contents

# S1. Theoretical calculations of Andreev bound states (ABSs)

## S1-1. Scattering matrix formalism

We utilize the scattering matrix formalism to explain the spectral features of our device. More precisely, we will use the description of the three-terminal graphene Josephson junction (*28, 48*) described by the random matrix theory (*17, 21, 54-56*). Our goal here is to consider the minimal, simplest model, which reproduces the spectral density of our experiments. Here, we assume that the time-reversal and spin-rotational symmetries are respected in the normal region because the graphene has negligible spin-orbit coupling. We will consider the short-junction limit because $L/\xi < 1$ where $L$ is the length scale of the normal region and $\xi$ is the coherence length.

We define $\Psi_{\text{in}}^{\text{e(h)}}$ as a vector of coefficients describing the wave incident on the normal region in the basis of electron(hole) modes. Similarly, $\Psi_{\text{out}}^{\text{e(h)}}$ is defined as a vector of coefficients describing the reflected and transmitted waves in the basis of electron(hole) modes. With a proper choice of the basis (*19, 56*) scattering in the normal region is described as follows (*28, 48*)

$$\begin{bmatrix} \Psi_{\text{out}}^{\text{e}} \\ \Psi_{\text{out}}^{\text{h}} \end{bmatrix} = \begin{bmatrix} s(E) & 0 \\ 0 & s^*(-E) \end{bmatrix} \otimes \sigma_0 \begin{bmatrix} \Psi_{\text{in}}^{\text{e}} \\ \Psi_{\text{in}}^{\text{h}} \end{bmatrix}$$

where $\sigma_i$ denotes Pauli matrices in spin space. In the same basis, the Andreev reflection is described by (*22, 28, 48, 54*)

$$\begin{bmatrix} \Psi_{\text{in}}^{\text{e}} \\ \Psi_{\text{in}}^{\text{h}} \end{bmatrix} = \alpha(E) \begin{bmatrix} 0 & r_A \\ -r_A^* & 0 \end{bmatrix} \otimes \sigma_0 \begin{bmatrix} \Psi_{\text{out}}^{\text{e}} \\ \Psi_{\text{out}}^{\text{h}} \end{bmatrix},$$

with $\alpha(E) = \sqrt{1 - \frac{E^2}{\Delta_{\text{ABS}}^2}} + i \frac{E}{\Delta_{\text{ABS}}}$. Here, $s(E)$ and $r_A$ are the appropriate unitary matrices, which describe scattering in the central normal region and Andreev reflection processes in the junction and have the dimension $3n \times 3n, n \in \mathbb{Z}^+$. Physically, $2n$ counts the number of incoming electronic modes per terminal, which must be even because of the spin degree of freedom. For simplicity, we further assume that all the terminals have the same number $2n$ of modes. As mentioned in the main text, the spectral density of states of our device is roughly consistent with that of the chaotic quantum dot (*54, 55*) described by the random matrix theory. Hence, we expect that the resulting spectral density is largely independent of the precise value of the modes per terminal, and we set $n = 2$ for our numerical simulation.

Now, we can write down the condition for the Andreev bound state (ABS) to satisfy

$$\begin{bmatrix} \Psi_{\text{in}}^{\text{e}} \\ \Psi_{\text{in}}^{\text{h}} \end{bmatrix} = \alpha(E) \begin{bmatrix} 0 & r_A \\ -r_A^* & 0 \end{bmatrix} \begin{bmatrix} s(E) & 0 \\ 0 & s^*(-E) \end{bmatrix} \otimes \sigma_0 \begin{bmatrix} \Psi_{\text{in}}^{\text{e}} \\ \Psi_{\text{in}}^{\text{h}} \end{bmatrix}.$$

In the short junction limit, the electron scattering matrix (in the normal region) can be approximated to be independent of energy (*17, 28, 48*), hence $s(E) \simeq s(-E) \simeq s(0) \equiv s$. After a few straightforward manipulations (*28*), we can find a much simpler equation for the ABS

$$(A \otimes \sigma_0)\Psi_{\text{in}}^{\text{e}} = \frac{|E|}{\Delta_{\text{ABS}}} e^{i\chi} \Psi_{\text{in}}^{\text{e}}$$

for some $\chi$. Here, $A = \frac{1}{2}(r_A^* s + s^T r_A^*)$. Therefore, the energy spectrum of the ABS is given by the absolute values of the eigenvalues of $A$ and their particle-hole partners. It is also useful to note that the time-reversal symmetry in the normal region constrains $s$ as an element of the circular orthogonal ensemble (COE) (*22, 53-56*), $s = s^T$. With this, we can further simplify $A$ as $A = \frac{1}{2}\{s, r_A^*\}$.

We will label the terminals R (on the right), L (on the left), and M (in between). Without loss of generality, the superconducting phases are given as $\varphi_{\text{R(L)}} \equiv 2\pi \frac{\Phi_{\text{R(L)}}}{\Phi_0}$ and $\varphi_{\text{M}} \equiv 0$, utilizing the gauge invariance. In an appropriate basis (*19, 56*), we can write $r_A$ as (*22, 28, 48, 54*)

$$r_A \equiv r_A(\varphi_R, \varphi_L) = \begin{bmatrix} ie^{-i\varphi_R} & 0 & 0 \\ 0 & ie^{i\varphi_L} & 0 \\ 0 & 0 & i \end{bmatrix} \otimes \mathbb{1}_n$$

where $\mathbb{1}_n$ is the $n \times n$ identity matrix. In the same basis, $s$ can be also written in a block matrix form (48, 54)

$$s = \begin{bmatrix} r_{RR} & t_{RL} & t_{MR} \\ t_{RL}^T & r_{LL} & t_{LM} \\ t_{MR}^T & t_{LM}^T & r_{MM} \end{bmatrix},$$

where $t_{ij}$ ($(i,j) = (R,L), (M,R), (L,M)$) is the $n \times n$ transmission matrix from $i$ to $j$, and $r_{ii}$ ($i = R, L, M$) is the $n \times n$ symmetric reflection matrix from $i$ to $i$.

**S1-1A. Linear dispersion relations:** In an alternative approach (17, 18), the ABS condition is provided as a determinant formula (48)

$$\det\left(\frac{\mathbb{1}_{6n}}{\alpha^2(E)} + r_A s^* r_A^* s \otimes \sigma_0\right) = 0.$$

For small $E$, $\frac{1}{\alpha^2(E)} \approx 1 - 2i\frac{E}{\Delta_{ABS}}$. Therefore, in the low-energy regime,

$$\det\left(E - i\frac{\Delta_{ABS}}{2}(\mathbb{1}_{6n} + r_A s^* r_A^* s \otimes \sigma_0)\right) = 0$$

holds.

Suppose that $E = 0$ at $(\varphi_R, \varphi_L) = (\phi_R, \phi_L)$. That is, $-1$ is an eigenvalue of $r_A s^* r_A^* s$. Let $|\Psi_0\rangle$ be an associated eigenvector. Due to the particle-hole symmetry, the eigenvalue $-1$ is indeed degenerate. Specifically, we find that $|\Psi_0'\rangle = r_A s^* |\Psi_0\rangle^*$ is also an associated eigenvector (17). Without loss of generality, $\langle\Psi_0|\Psi_0\rangle = \langle\Psi_0'|\Psi_0'\rangle = 1$ and $\langle\Psi_0|\Psi_0'\rangle = \langle\Psi_0'|\Psi_0\rangle = 0$.

We consider a small perturbation in $(\varphi_R, \varphi_L)$-space. For any $|\delta\phi_R|, |\delta\phi_L| \ll 1$, let $(\phi_R + \delta\phi_R, \phi_L + \delta\phi_L)$ be the point of our interest. Up to first order terms,

$$r_A s^* r_A^* s \approx r_{A,0} s^* r_{A,0}^* s - i\delta\phi r_{A,0} s^* r_{A,0}^* s + i r_{A,0} s^* r_{A,0}^* s(s^* \delta\phi s).$$

Here,

$$r_A = r_A(\phi_R + \delta\phi_R, \phi_L + \delta\phi_L),$$
$$r_{A,0} = r_A(\phi_R, \phi_L),$$
$$\delta\phi = \text{diag}(\delta\phi_R, -\delta\phi_L, 0) \otimes \mathbb{1}_n.$$

We project $r_A s^* r_A^* s$ onto the subspace whose basis is $\{|\Psi_0\rangle, |\Psi_0'\rangle\}$ (17, 18):

$$r_A s^* r_A^* s \mapsto -\rho_0 - i\begin{bmatrix} \langle\Psi_0|s^*\delta\phi s - \delta\phi|\Psi_0\rangle & \langle\Psi_0|s^*\delta\phi s - \delta\phi|\Psi_0'\rangle \\ \langle\Psi_0'|s^*\delta\phi s - \delta\phi|\Psi_0\rangle & \langle\Psi_0'|s^*\delta\phi s - \delta\phi|\Psi_0'\rangle \end{bmatrix},$$

while Pauli matrices in the subspace are denoted as $\rho_i$. Then,

$$\det\left(E - \frac{\Delta_{ABS}}{2}\begin{bmatrix} \langle\Psi_0|s^*\delta\phi s - \delta\phi|\Psi_0\rangle & \langle\Psi_0|s^*\delta\phi s - \delta\phi|\Psi_0'\rangle \\ \langle\Psi_0'|s^*\delta\phi s - \delta\phi|\Psi_0\rangle & \langle\Psi_0'|s^*\delta\phi s - \delta\phi|\Psi_0'\rangle \end{bmatrix} \otimes \sigma_0\right) = 0.$$

After a few manipulations, we find the low-energy effective Hamiltonian as

$$H_{\text{eff}} = \sum_{i=1}^{3} h_i \rho_i \otimes \sigma_0$$

with $h_1 - ih_2 = -\Delta_{ABS}\langle\Psi_0|\delta\phi|\Psi_0'\rangle$ and $h_3 = \frac{\Delta_{ABS}}{2}(\langle\Psi_0'|\delta\phi|\Psi_0'\rangle - \langle\Psi_0|\delta\phi|\Psi_0\rangle) \in \mathbb{R}$, thus $E = \pm\sqrt{\sum_{i=1}^{3} h_i^2}$.

Since $h_i$ are linear forms in $\delta\phi_R$ and $\delta\phi_L$, $E$ is a square root of a positive semidefinite quadratic form, up to sign. That is, in $(\delta\phi_R, \delta\phi_L, E)$-space, the energy surface is either an elliptic cone or two intersecting planes. Thus, near any zero-crossing point, low-energy bands are to be linearly dispersive.

**S1-1B. Example:** As an illustration, let us consider an example. This model will be later shown to saturate the bound for the energy spectrum. In this example, we will set $n = 2$, the smallest allowed value of $n$. We will also choose the most symmetric forms of the scattering and reflection matrices

$$r_{RR} = r_{LL} = r_{MM} = \begin{bmatrix} \gamma & 0 \\ 0 & -\gamma \end{bmatrix},$$

and

$$t_{RL} = t_{MR}^T = t_{LM} = \begin{bmatrix} \frac{1-\gamma}{2} & -\frac{\sqrt{(1-\gamma)(1+3\gamma)}}{2} \\ \frac{\sqrt{(1-\gamma)(1+3\gamma)}}{2} & -\frac{1-\gamma}{2} \end{bmatrix}$$

with a tunable value $0 \leq \gamma \leq 1$. It is easy to confirm that the above scattering matrix satisfies the time-reversal and spin-rotation symmetries, and thus it belongs to the COE, as required. Here, $\gamma$ represents the degree of 'reflection' for $s$. For a given value of $\gamma$, we can easily evaluate the spectrum of the ABS. For example, the ABS bands at $\gamma = 0.25$ can be found in Fig. 1**B**, which exhibits the nodal lines for $0 \leq \gamma < \frac{1}{3}$ in the space of $(\varphi_L, \varphi_R)$. On the other hand, Nodal points appear for $\gamma = \frac{1}{3}$, and the spectrum is fully gapped for $\gamma > 1/3$.

## S1-2. Infimum of non-negative ABS energies

As verified in Ref. 28, the smallest modulus of weighted means of eigenvalues of $r_A$ is the non-trivial lower bound of all possible ABS energy levels (for their absolute values). This lower bound is saturated by our example above and is well respected in experimental data.

We will notate $|\Psi\rangle \equiv \Psi_{in}^e$, $|\Psi'\rangle \equiv (s \otimes \sigma_0)|\Psi\rangle$. Then, trivially $\langle\Psi|\Psi\rangle = \langle\Psi'|\Psi'\rangle = 1$. With this, the ABS condition is written as (*28*)

$$\langle\Psi|r_A^* \otimes \sigma_0|\Psi\rangle + \langle\Psi'|r_A^* \otimes \sigma_0|\Psi'\rangle = \frac{2|E|}{\Delta_{ABS}} e^{i\chi}\langle\Psi'|\Psi\rangle.$$

Consequently, we get (*28*)

$$\frac{|E|}{\Delta_{ABS}} \geq \frac{1}{2}|\langle\Psi|r_A^* \otimes \sigma_0|\Psi\rangle + \langle\Psi'|r_A^* \otimes \sigma_0|\Psi'\rangle|,$$

implying that $\frac{|E|}{\Delta_{ABS}}$ is no less than the absolute value of any weighted averages of eigenvalues of $r_A^*$, or equivalently $r_A$.

For the three-terminal Josephson junction, we can rigorously determine the infimum of non-negative ABS energies. For the given values of $\varphi_R$ and $\varphi_L$, let $S(\varphi_R, \varphi_L)$ be the set of all possible ABS energy levels. Then,

$$\inf\left\{\frac{|E|}{\Delta_{\text{ABS}}}\,\middle|\, E \in S(\varphi_R, \varphi_L)\right\}$$

$$= \begin{cases} 0 & \text{if } m_1 \leq \varphi_R \bmod 2\pi \leq m_2 \text{ for } \{m_1, m_2\} = \{\pi, (\pi - \varphi_L) \bmod 2\pi\} \\ \min\left\{\left|\cos\left(\frac{\varphi_R}{2}\right)\right|, \left|\cos\left(\frac{\varphi_L}{2}\right)\right|, \left|\cos\left(\frac{\varphi_R + \varphi_L}{2}\right)\right|\right\} & \text{otherwise.} \end{cases}$$

Here, the modulo operation maps $\mathbb{R}$ onto $[0, 2\pi)$. We verify that it is indeed the greatest of all lower bounds by analytically determining the energy spectrum for $\gamma = 0$ and zero energy crossing points for all $0 \leq \gamma \leq \frac{1}{3}$ for the above example [see Fig. S1].

**S1-2A. Linear dispersion relations:** Suppose that $(\phi_R, \phi_L)$ is in the boundary of the 'gapless region' $R_0$. To be rigorous, $R_0$ is the set of points in $(\varphi_R, \varphi_L)$-space where the infimum of non-negative ABS energies is zero. It is straightforward that $\phi_R \bmod 2\pi = \pi$, $\phi_L \bmod 2\pi = \pi$, or $(\phi_R + \phi_L) \bmod 2\pi = \pi$.

Consider a small variation $\delta\phi = (\delta\phi_R, \delta\phi_L)$. By small-angle approximation,

$$\inf\left\{\frac{|E|}{\Delta_{\text{ABS}}}\,\middle|\, E \in S(\phi_R + \delta\phi_R, \phi_L + \delta\phi_L)\right\} \in \left\{\frac{|\delta\phi_R|}{2}, \frac{|\delta\phi_L|}{2}, \frac{|\delta\phi_R + \delta\phi_L|}{2}\right\}$$

unless the infimum is zero. Therefore, near the boundary of the 'gapless region', the infimum of non-negative ABS energies is either zero or linear in $\delta\phi$.

## S2. Numerical simulations

### S2-1. ABS physics in three-terminal Josephson junction

We present the details for numerical calculation of DOS of Andreev bound states in three-terminal Josephson junction within the random matrix approach (*28, 53, 54*). Based on the universality of the random matrix theory, we expect that the precise value of $n$ is largely irrelevant to the resulting DOS. Hence, we fix $n = 2$ for simplicity. Hence, $s$ and $r_A$ are given as $6 \times 6$ matrices. First, we sample the 50,000 random $s$ matrices from $\text{COE}(6)$. Here, the sampling is constrained to model the high junction transparency [see Supplementary section S2-1A], which is consistent with experimental data. Next, for each $s$, $\frac{\Phi_R}{\Phi_0}$ and $\frac{\Phi_L}{\Phi_0}$, we construct the matrix $A$. Finally, we numerically calculate its eigenvalues to find the ABS energy spectrum. Then, we calculate the DOS from the spectrum.

**S2-1A. Junction transparency:** We present a phenomenological approach to systematically control the junction transparencies in scattering matrix samples. In this work, we use the Frobenius norm as an indicator of the degrees of 'transmission' and 'reflection' among the three terminals. The Frobenius norm of a matrix $W_{ij}$ is given by

$$\|W\|_F = \sqrt{\sum_{i,j} |W_{ij}|^2} = \sqrt{\sum_i \lambda_i^2},$$

where $\lambda_i$ is the eigenvalue of $W$. When $W$ is a reflection matrix, $\|W\|_F$ is equivalent to $\sqrt{\sum_a R_a}$, where $R_a$ is the reflection probability of the $a$-th eigenmode for $W$. Hence, the above norm can parametrize how good the reflection is in each terminal.

Since the junction transparency is related to the sum of reflection probabilities over terminals, the transparency can be tuned by adjusting the value $\| r_{RR} \|_F^2 + \| r_{LL} \|_F^2 + \| r_{MM} \|_F^2$. To be specific, we constrain the sampling of $s$ matrices with the inequality

$$\| r_{RR} \|_F^2 + \| r_{LL} \|_F^2 + \| r_{MM} \|_F^2 < 1,$$

representing the case where the junction is highly transparent, consistent with our experiment. This distribution of reflection probabilities affects the spectral feature of the junction.

**S2-1B. Anisotropic connectivity between terminals:** To model the anisotropic connectivity between terminals, we adopt a similar phenomenological method to measure the degree of connectivity in the sampling procedure.

When $W$ is a transmission matrix, $\| W \|_F$ is equivalent to $\sqrt{\sum_a T_a}$, where $T_a$ is the transmission probability of the $a$-th eigenmode for $W$. That is, the Frobenius norm can also parametrize how good the transmission is between two terminals.

Therefore, in the case where M is relatively less connected to the others, we impose the scattering matrices $s$ to satisfy

$$\max\{\| t_{MR} \|_F, \| t_{LM} \|_F\} < 0.8 \| t_{RL} \|_F.$$

It represents the case where the transmission between R and L is much larger than the transmission between R and M or between L and M. Under such condition, we obtain Fig. 4C, which is largely consistent with the experimental data.

In fact, the above is a particular case of a general relation

$$\max\{\| t_{MR} \|_F, \| t_{LM} \|_F\} < c_M \| t_{RL} \|_F,$$

where the degree of anisotropy can be tuned by the parameter $c_M \in (0, \infty)$. When $c_M \to \infty$, the junction becomes isotropic. As $c_M$ decreases, the system becomes more anisotropic. Such behavior is depicted in Fig. S7.

Similarly, to simulate the case where R is not well-connected to the others, we choose $s$ matrices to meet the condition

$$\max\{\|\| t_{RL} \|_F\ t_{MR} \|_F\} < 0.8 \| t_{LM} \|_F.$$

It depicts the situation where the transmission between L and M being far larger than the transmission between R and L or between M and R. Under such condition, we obtain Fig. S8, which is also largely consistent with the experimental data.

Finally, we note that

$$\| t_{ij} \|_F^2 = \frac{1}{2}\left(n - \| r_{ii} \|_F^2 - \| r_{jj} \|_F^2 + \| r_{kk} \|_F^2\right), \quad \| r_{ii} \|_F^2 = n - \| t_{ki} \|_F^2 - \| t_{ij} \|_F^2,$$

where $i \neq j$, $j \neq k$. Also, the following is true:

$$\max\{\| t_{MR} \|_F, \| t_{LM} \|_F\} < \| t_{RL} \|_F \Leftrightarrow \max\{\| r_{RR} \|_F, \| r_{LL} \|_F\} < \| r_{MM} \|_F.$$

Hence, we find an intuitive conclusion that the good connectivity between $R$ and $L$ (i.e., $\| t_{RL} \|_F$ is the largest) is equivalent to the fact that M is the least transparent (i.e., $\| r_{MM} \|_F$ is the largest). The result and its physical implication are both consistent with those from the quasiclassical Green-function formalism (*29*).

## S2-2. Theoretical estimation of differential conductance

From the DOS of ABSs, we can next theoretically calculate the differential conductance by numerical integration [see Figs. 2, 3]. The differential conductance is a weighted sum of integral transforms given by (*16*)

$$\frac{dI}{dV} \sim \int \text{DOS}_{\text{ABS}}(E)\, K(E,V)\, dE + w \int \text{DOS}_{\text{ind}}(E)\, K(E,V)\, dE$$

by the kernel

$$K(E,V) = \frac{d\text{DOS}_{\text{probe}}(E+eV)}{dV}(f(E) - f(E+eV)) - \text{DOS}_{\text{probe}}(E+eV)\frac{df(E+eV)}{dV}.$$

Here, $\text{DOS}_{\text{ABS}}(E)$ is the density of ABSs, $\text{DOS}_{\text{ind}}(E)$ is the additional DOS due to the proximity-induced superconductivity from superconductors to graphene, $\text{DOS}_{\text{probe}}(E)$ is the DOS of the probe, and $f(E)$ is the Fermi-Dirac distribution. To be specific,

$$\text{DOS}_{\text{probe}}(E) = \left| \text{Re}\left( \frac{E + i\gamma_{\text{probe}}\Delta_{\text{Al}}}{\sqrt{(E + i\gamma_{\text{probe}}\Delta_{\text{Al}})^2 - \Delta_{\text{Al}}^2}} \right) \right|,$$

and

$$\text{DOS}_{\text{ind}}(E) = \left| \text{Re}\left( \frac{E + i\gamma_{\text{ind}}\Delta_{\text{ABS}}}{\sqrt{(E + i\gamma_{\text{ind}}\Delta_{\text{ABS}})^2 - \Delta_{\text{ABS}}^2}} \right) \right|,$$

with the Dynes parameters $\gamma_{\text{probe}}$ and $\gamma_{\text{ind}}$ (*16*). The ratio between tunneling amplitudes from the probe to the proximity-induced superconducting states and to ABSs is considered by the weight $w$. Furthermore, since the temperature is sufficiently low, we set

$$f(E) = \frac{1}{\exp(E/k_B T) + 1} \approx \Theta(-E),$$

where $\Theta$ is the Heaviside step function. Also, we set $\Delta_{\text{ABS}} = 0.155$ meV and $\Delta_{\text{Al}} = 0.19$ meV [see Supplementary section S5]. Rest of the parameters are chosen so that the theoretical estimations of differential conductance match the experimental data: $\gamma_{\text{probe}} = 0.18$, $\gamma_{\text{ind}} = 0.1$, and $w = 0.1$.

## S3. Phase and temperature dependence of Andreev bound states in single-loop SQUID

Figure S3A shows the differential conductance, $dI/dV$ measured at a bath temperature of 20 mK as a function of bias voltage, $V_B$ and superconducting phase difference, $\varphi$. The peaks in $dI/dV$ appear at the bias voltage $V(\varphi) = [\pm\Delta_{\text{Al}} + E^{\pm}(\varphi)]/e$, where $E^{\pm}(\varphi) = \Delta_{\text{ABS}}\sqrt{(1 - \tau \sin^2(\varphi/2))}$. Fitting the oscillations of the peak position as a function of $\varphi$, we obtain contact transparency, $\tau = 0.97$, and $\Delta_{\text{ABS}} = 0.155$ meV for the upper ($E^+$) and lower ($E^-$) bands of ABS, with $E^{\pm}(\varphi = 0) = \pm\Delta_{ABS}$. Here, $\Delta_{\text{Al}}$ is the superconducting gap of the Al tunnel probe, which is 0.175 meV, cross-verified later in the temperature dependence. The near-perfect transparency, indicated by the high $\tau$, reveals highly transparent superconducting contacts to graphene. It's also worth noting that the coherence length, $\xi = \frac{\hbar v_F}{\Delta_{\text{ABS}}} = 4.2$ μm is longer than the channel length, $L=1$ μm, indicating that the GJJ operates in the short junction limit. The critical temperature of ohmic-contacted Al electrodes ($T_{\text{c,ABS}}$), where $Δ_{ABS}$ vanishes, and tunnel-contacted Al electrodes ($T_{c,Al}$), where $Δ_{Al}$ vanishes, possess different values and thus, exhibit distinct slopes in response to temperature. Consequently, the temperature-dependent d$I$/d$V$ peak position at $φ$=0 (circle symbol), depicted in Fig. S3B, is fitted to $V(T) = [Δ_{Al}(T) + Δ_{ABS}(T)]/e$(Black solid line), using two fitting parameters, $T_{c,ABS}$ and $T_{c,Al}$. Here, $Δ_{ABS}(T)$ follows the BCS temperature dependence with $Δ_{ABS}(T=0)$= 0.155 meV and $T_{c,ABS}$=1.02 K, while $Δ_{Al}(T)$ follows BCS temperature dependence with $Δ_{Al}(T=0)$= 0.175 meV and $T_{c,Al}$=1.15 K.

## S4. Characterization of graphene Josephson junction through transport measurements

Figure S4A presents the measurements of normal-state resistance, $R_N$ at a temperature of 0.9 K, across various values of gate voltage, $V_G$. The resistance asymmetry between the negative and positive sides of $V_G$ arises from electron doping in the graphene layer near the electrode contact, caused by a mismatch in the Fermi level. Consequently, this results in a junction resistance at positive $V_G$ that is significantly lower than that on the negative side, where a less transmitting n-p-n junction is formed. The observation of Fabry-Perot oscillation at negative $V_G$ (see also Fig. S4C), attributable to the interference of carriers reflected at the two p-n boundaries, signifies the ballistic transport in the graphene region. Therefore, the comparison of $R_N$ with the ballistic-limit value of $R_Q = (h/4e^2)$ (1/$N$) allows for the determination of the contact transparency, $τ = R_Q/R_N$ as shown in Fig. S4B. Here, $N = 2W/λ_F$ represents the number of carrier propagation modes for the channel width ($W$ = 1 μm) and the Fermi wavelength, $λ_F$. Additionally, the cavity length was estimated from the resonance condition 2$L_{cavity}$ = m$λ_F$, where $L_{cavity}$ is the cavity length (the region of graphene unaffected by the electrode doping), and m is an integer. This was derived from the $V_G$ difference between the adjacent peaks as depicted in Fig. S4C. The channel length, approximated by the SEM, also gives $L$~1 μm, which is consistent with the length ascertained from the oscillation period of $V_G$. Figure S4D shows the critical current, $I_c$, of the junction as a function of the $V_G$. The $I_c$ ($V_G$) also exhibited asymmetry around the charge neutrality point ($V_{Dirac}$=-3 V), similar to the asymmetry observed in $R_N$. In the highly transmitting n-doped region, $I_c$ exceeds approximately 200 nA, while it drops to roughly 20 nA in the p-doped region. These observations provide direct evidence of ballistic Josephson coupling in our devices. This coupling is facilitated by the coherent transport of ballistic quasi-particles through the ABS channel in graphene.

## S5. Phase and temperature dependence of ABSs in double-loop SQUID

Figure S5A presents the scenario of maximum d$I$/d$V$, where the DOS peak of the tunnel probe aligns with the ABS at an energy derived from bias voltage, $eV_B = (Δ_{Al}+Δ_{ABS})$. Here, $Δ_{Al}$ is the superconducting gap of the Al tunnel probe. The sharp peak in the DOS of the tunnel probe near $Δ_{Al}$ facilitates high energy resolution in tunneling spectroscopy. The left panel in Fig. S5B shows the characteristic behavior of d$I$/d$V$ as a function of bias voltage, $V_B$, along the phase-space line corresponding to red colored dashed arrow in Fig. 2**A** in the main text (This data corresponds to full energy spectrum of Fig. 3**A** in the main text). The right panel in Fig. S5B shows line cuts at $φ = 0$ and $π$. These line cuts reveal peaks in the d$I$/d$V$ attributable to the upper (E$^+$) and lower (E$^-$) bands of the Andreev band. Through the peak position at $φ = 0$, we determine the induced ABS gap, $Δ_{ABS}$= 0.155 meV for double loop SQUID device1. Here, $Δ_{Al}$ is verified to be 0.19 meV later in the temperature dependence analysis. It is noteworthy that the coherence length, $ξ = \frac{ℏv_F}{Δ_{ABS}}$=4.2 μm, exceeds the maximum channel length, $L$=1 μm of the 3TJJ, indicating that the GJJ operates in the short junction limit. The critical temperature of ohmic-contacted Al electrodes ($T_{c,ABS}$), where $\Delta_{ABS}$ vanishes, and tunnel-contacted Al electrodes ($T_{c,Al}$), where $\Delta_{Al}$ vanishes, possess different values and thus, exhibit distinct slopes in response to temperature. Consequently, the temperature-dependent d$I$/d$V$ peak position at $\varphi = 0$ (circle symbol), depicted in Fig. S5C, is fitted to $V(T) = [\Delta_{Al}(T) + \Delta_{ABS}(T)]/e$ (Black solid line), using two fitting parameters, $T_{c,ABS}$ and $T_{c,Al}$. Here, $\Delta_{ABS}(T)$ follows the BCS temperature dependence with $\Delta_{ABS}(T = 0)= 0.155$ meV and $T_{c,ABS}$=1.02 K, while $\Delta_{Al}(T)$ follows BCS temperature dependence with $\Delta_{Al}(T = 0)= 0.19$ meV and $T_{c,Al}$=1.25 K. As anticipated, the effects of the Fermi distribution at non-zero temperatures become evident through a broadening of the d$I$/d$V$ and a gradual diminishing of its main features as bath temperature $T_0$ increases. In particular, the topological gapped regions are still clearly observable up to approximately 0.9 K.

# S6. Characteristics of zero-energy crossing in iso-energy surface at different gate voltages

In ballistic graphene, the number of carrier propagation modes, $N$, can be determined using the equation $N \approx W k_F/\pi$, where $k_F = \sqrt{\pi c_g |V_g|/e}$ is the Fermi wavevector, $c_g$ represents the gate capacitance per unit square (fF/μm$^2$), and $W$ is the width of the contacted graphene. In the short junction limit, the number of ABSs is equal to the number of conduction modes in the normal region. Therefore, by varying the gate voltage, $V_G$, the Josephson coupling can be adjusted. Figure S6 illustrates the characteristics of zero-energy crossing within the iso-energy surface, dependent on $V_G$. $V_G$ is tuned from hole-type (panels A-B) through near the $V_{Dirac}$ (panels C-D) to electron-type (panels E-F). As expected, the greater the carrier density, the stronger the zero-energy crossing feature is modulated with phase, owing to a larger Josephson coupling, which agrees with the transport measurements of GJJ (see Fig. S4D). In contrast, at low carrier density, close to the $V_{Dirac}$ (Fig. S6C), the phase modulation of the d$I$/d$V$ is very weak. Noticeably, in the hole-type range, the phase modulation of d$I$/d$V$ is relatively weaker than in the electron-type range. This effect is attributed to decoherence, which causes the formation of an n-p-n junction in graphene when its $V_G$ is hole-doped.

# S7. Anisotropy of M terminal connectivity

In Fig. S7, we examine the anisotropic connectivity dependence of the DOS at zero-energy (panel A-E) and in the energy spectrum (panel F-J). We use the Frobenius norm of matrices as a measure to effectively quantify the degrees of 'transmission' and 'reflection' among the terminals, as elaborated in Supplementary section S2-2. Each panel represents a different degree of connectivity for the M terminal. Progressing from left to right, the connectivity of the M terminal to the other terminals (or the transmission between M and R, L terminals) gets weaker compared to the connectivity between R and L terminals (or the transmission between R and L terminals). In Figs. S7A-E, we observe that decreasing connectivity of the M terminal leads to an anisotropic density of the gapless states in the phase space. In the isotropic case (panel A), we obtain an isotropic, triangular shape of the gapless state (red region). These gapless regions are completely enclosed by Kagome-like closed loops in the 2D phase space. As the connectivity of the M terminal decreases, these closed loops of gapless regions begin to disconnect from one another, except in the principal direction where $\varphi_L = -\varphi_R$. In the case where the M terminal is almost disconnected (panels D, E), a two-terminal like Josephson junction is governed by the total phase $\varphi_L + \varphi_R$. Figure S7F-J further illustrates how the anisotropy of the M terminal's connectivity influences the evolution of the full energy spectrum of the $dI/dV$, in the direction, defined by $\varphi$ $(= \varphi_L = \varphi_R)$. The most notable feature is the extensive deduction of the $dI/dV$ that spreads from the low energy to the high energy around $\varphi = \pm\pi$. This deduction of the $dI/dV$ culminates at the black dashed line, which has a period equal to the periodicity of the single ABS. This tendency signifies a crossover from a three-terminal ABS to a two-terminal ABS as the connectivity of the M terminal weakens.

## S8. Anisotropy of R terminal connectivity for device 2

Figure S8A-F provides a tomographic representation of the experimental results obtained from device2, with measurements taken at a fixed $V_G = 20$ V. Here, the $dI/dV$ is plotted as a function of $(\varphi_L, \varphi_R)$ for various $(eV_B - \Delta_{Al})/\Delta_{ABS}$. The superconducting gap of the Al tunnel probe is denoted by $\Delta_{Al} = 0.19$ meV, and the induced ABS gap is $\Delta_{ABS} = 0.155$ meV. For comparative insights, Figs. S8G-L displays our theoretically-calculated $dI/dV$ with anisotropic connectivity of the R terminal. As the connectivity of the R terminal diminishes, the ABS resonance along the $\varphi_R$ direction in 2D phase space becomes pronounced, whereas the ABS resonance between the horizontal axes becomes weaker.

## S9. Estimation of electron temperature depending on the flux gate current

The flux gate line, fabricated from an Al superconductor, remains thermally neutral. Yet, the incorporated RC filter within the measurement system linked to this line is a source of heat. This culminates in an elevated temperature of the dilution refrigerator. Recognizing potential discrepancies between the heightened dilution and device temperatures, we employed the GJJ as a device temperature sensor. We estimate the device temperature, $T_{device}$ depending on the flux current by comparing the $I_c$ dependence on bath temperature, $T_0$ and the flux current. At the peak of $dV/dI$, highlighted by red circles, signifies the critical current, $I_c$. Figure S10A shows the $T_0$ dependence of $I_c$ at the $V_G = 20\,V$. Similar to the $T_0$ dependence of the peak position (red circles) $I_c$ decreases as the flux current increases in Fig. S10B. This implies that the flux current increases the $T_{device}$. By matching the $I_c$ values, we can estimate the flux current dependence of the $T_{device}$ as shown in Fig. S10C. Converting flux current to normalised magnetic flux, applying $\Phi_R/\Phi_0$ raises the $T_{device}$ by approximately 0.3 K. As can be seen from the results in Fig. S5C, measured without the flux gate, there is no problem observing the transition feature in the range of $\pm\Phi/\Phi_0$ when measured with the flux gate.

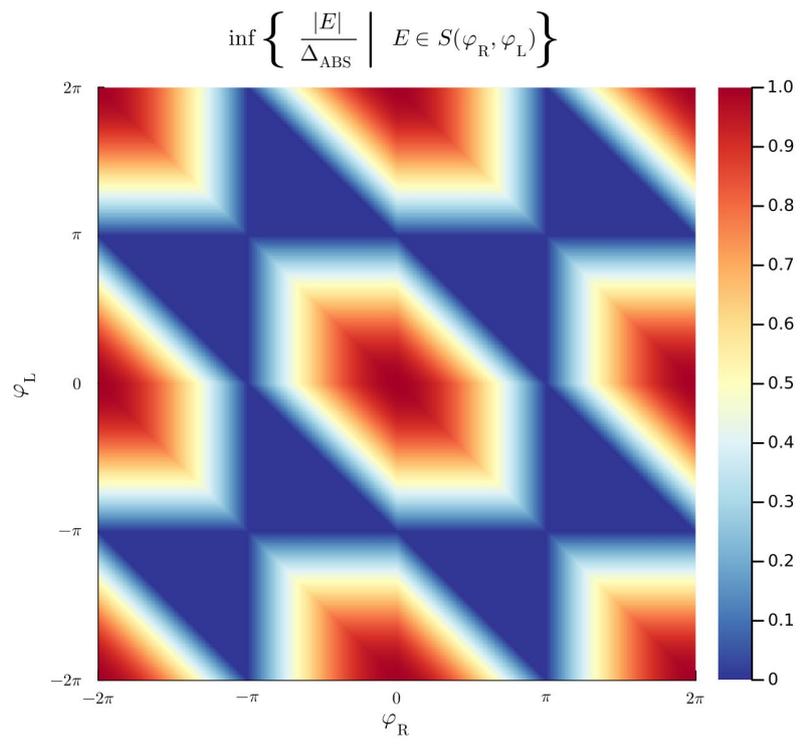

**Fig. S1. Infimum of all possible ABS energy levels in absolute values.**

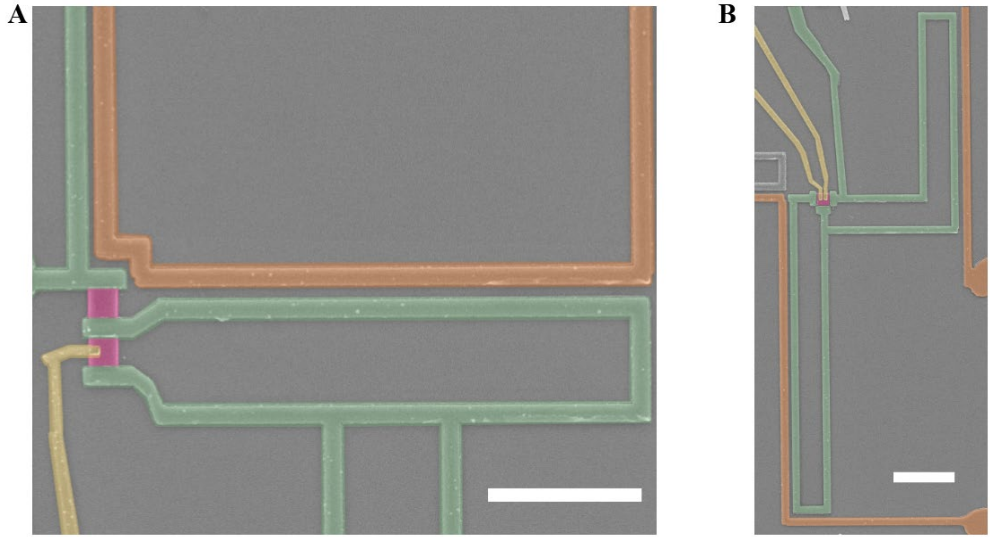

**Fig. S2. False-colored Scanning Electron Microscope (SEM) image.** (**A**) False-colored SEM image of a single-loop SQUID and Graphene Josephson junction (GJJ) situated between two neighboring SQUID loops, used for extracting a transport characteristic of the device. (**B**) False-colored SEM image of a double-loop SQUID denoted as device2. Aluminum electrodes (green) are tunneling contacted and Al electrodes (yellow) are ohmic-contacted to graphene (magenta). The magnetic field induced by the flux gate (orange) threading a superconducting loop controls $\varphi$ and modulates the Andreev state energy. The scale bar is 5 μm.

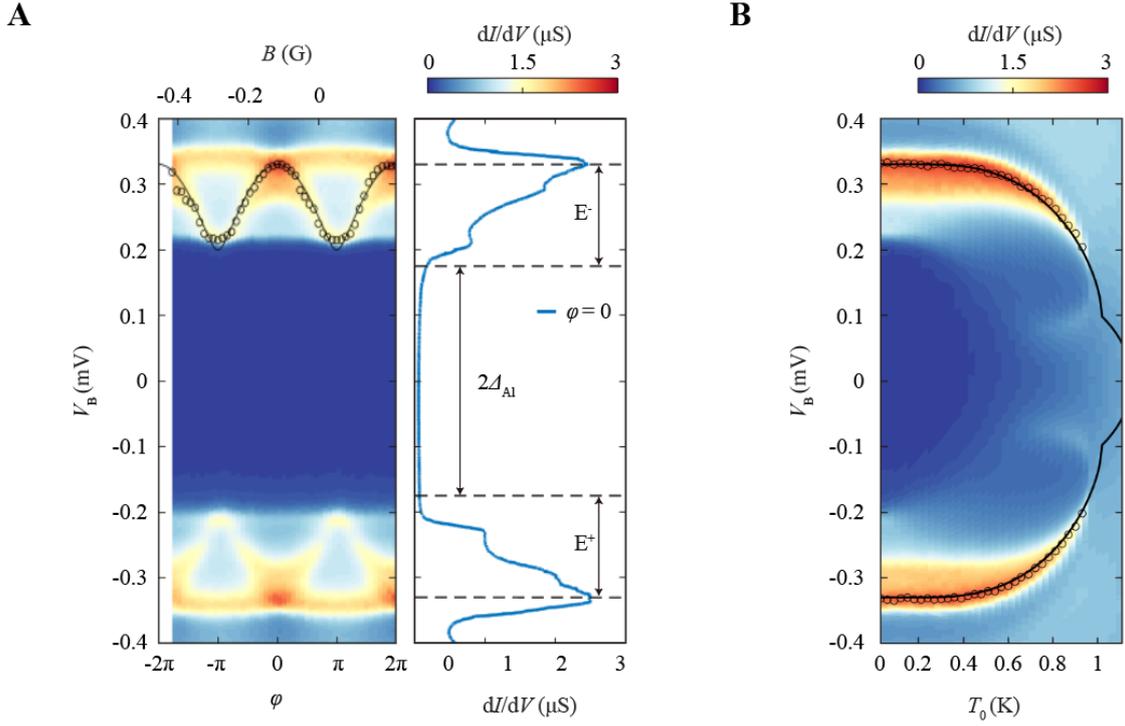

**Fig. S3. Phase and Temperature Dependence of Andreev bound states in Single-loop SQUID.** (**A**) Left panel shows a colour-coded plot of differential conductance, d$I$/d$V$ as a function of bias voltage, $V_B$ and superconducting phase difference, $\varphi$ of the single-loop SQUID. Black solid lines represent theoretical fitting in a short junction limit. The right panel shows a line cut at $T$ = 20 mK at $\varphi$=0. The line cut shows d$I$/d$V$ peaks due to the upper ($E^+$) and lower ($E^-$) bands of ABS. (**B**) Colour-coded plot of d$I$/d$V$ as a function of $V_B$ and temperature $T$ at $\varphi$=0. The value of the bias voltage at the peak of d$I$/d$V$ (black circle), decreases as the temperature increases. This trend is fitted to Bardeen-Cooper-Schrieffer (BCS) theory (black solid line).

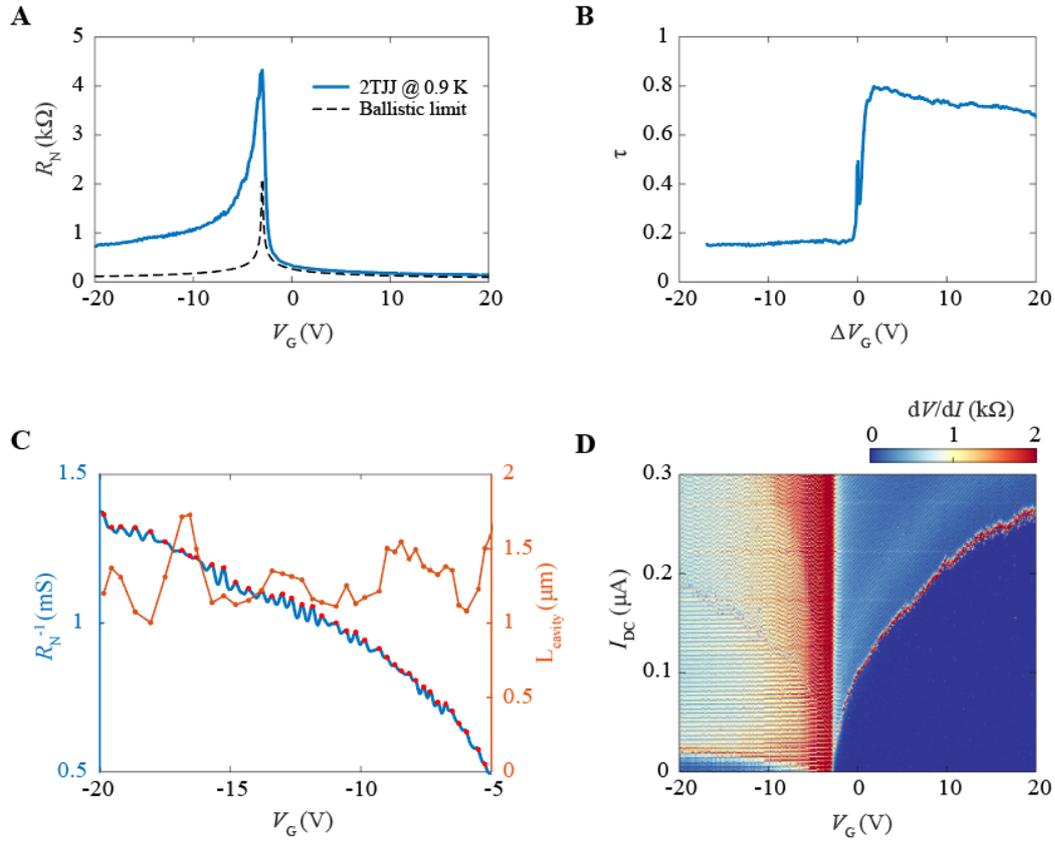

**Fig. S4. Characterization of GJJ through Transport measurements.** (**A**) Normal-state junction resistance, $R_N$ of GJJ measured at 0.9 K for varying backgate voltage, $V_G$. The black dashed line corresponds to the ballistic limit with perfect transparency ($\tau=1$) (**B**) Contact transparency, $\tau$ is plotted as a function of the $V_G$. (**C**) Fabry-Perot oscillation of the normal-state conductance $(R_N)^{-1}$ of graphene junction, which indicates ballistic pair transport. The length of the Fabry-Perot cavity was estimated from the resonance condition marked red dot. (**D**) $dV/dI$ map of GJJ at 20 mK as a function of the bias current, $I_{DC}$ and $V_G$, where the boundary of the red region represents $I_c(V_G)$ of the junction.

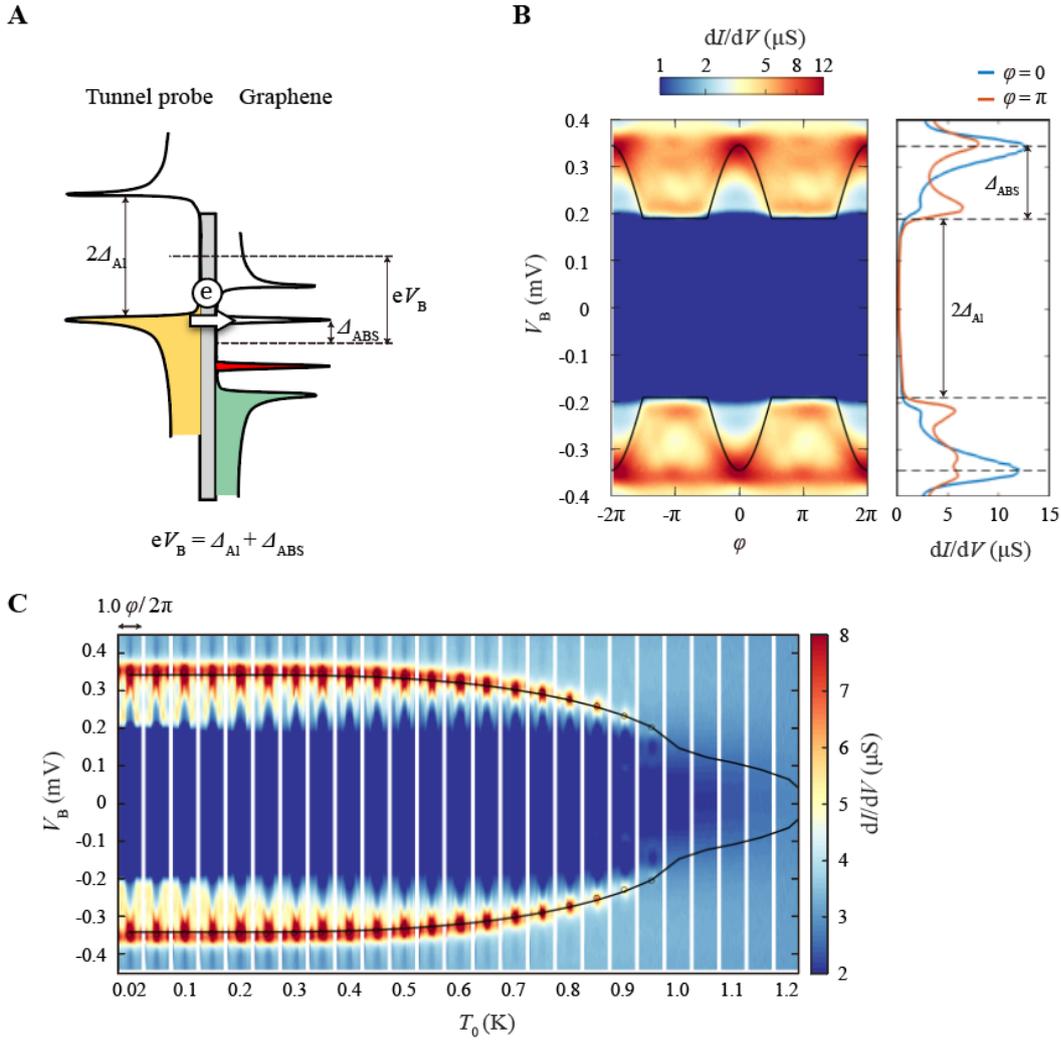

**Fig. S5. Phase and Temperature Dependence of ABSs in Double-loop SQUID (Device1).**

(**A**) The schematic illustrates the tunneling process when a bias voltage $V_B<0$, is applied between the tunnel probe and graphene. Specific bias voltage condition results in the alignment of the occupied DOS peak of the tunnel probe with the vacant DOS peak of the ABS. The value $eV_B-\Delta_{Al}$ represents the induced ABS gap in graphene, denoted as $\Delta_{ABS}$. (**B**) The left panel shows the full energy spectrum of Fig. 3A in the main text. This spectrum is plotted along the phase-space line cut, defined by equal phase $\varphi\ (=\varphi_L=\varphi_R)$. The black solid lines represent the analytical lower bound of the Andreev band, which refers to a theoretical prediction for the minimum energy gap within a short junction. The right panel shows a line cut at $\varphi=0$ and $\pi$. These line cuts reveal peaks in the d$I$/d$V$ attributable to the upper ($E^+$) and lower ($E^-$) bands of the Andreev band. (**C**) Colour-coded plot showing the data from (**B**), represented at various bath temperatures, $T_0$, within the range of -0.5 to 0.5 $\varphi/2\pi$. The value of the $V_B$ at the peak of d$I$/d$V$ (black circle) at $\varphi=0$, decreases as the temperature increases. This trend is fitted to Bardeen-Cooper-Schrieffer (BCS) theory (black solid line).

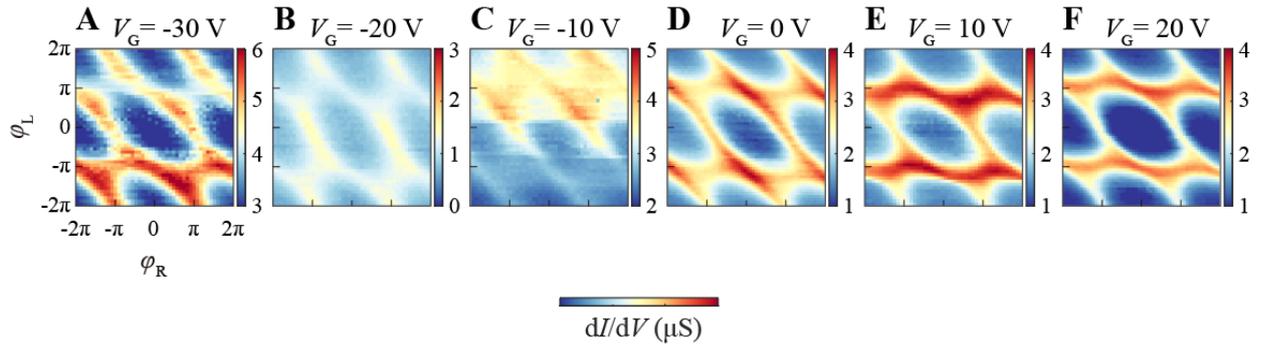

**Fig. S6. Characteristics of Zero-Energy Crossing in Iso-Energy Surface at Different Gate Voltage.** (A-F) Experimentally measurement of differential tunneling conductance, $dI/dV$, as a function of $(\varphi_L, \varphi_R)$, for zero-energy condition, $(eV_B - \Delta_{Al})/\Delta_{ABS} = 0$, across different gate voltage values.

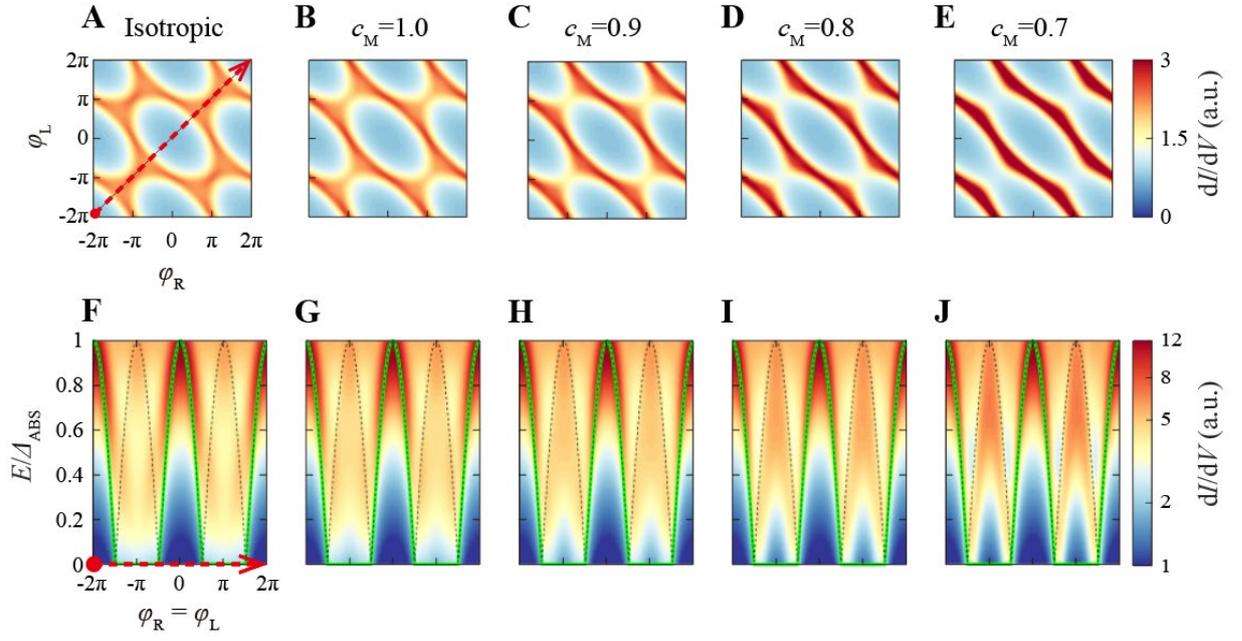

**Fig. S7. Anisotropy of M Terminal Connectivity (A-E)** Theoretically calculated $dI/dV$ as a function of $(\varphi_L, \varphi_R)$ at zero-energy. Each panel corresponds to a different connectivity of the M terminal. The red dashed arrow indicates where the fluxes in the two loops are equal, represented as $\varphi\ (=\varphi_L=\varphi_R)$. **(F-J)** Theoretically calculated $dI/dV$ are presented as a function of normalised energy, $E/\Delta_{ABS}$, and bias voltage, $V_B$. The black dashed line represents the analytic lower bound of a two-terminal ABS when the connectivity of the M terminal is weakened.

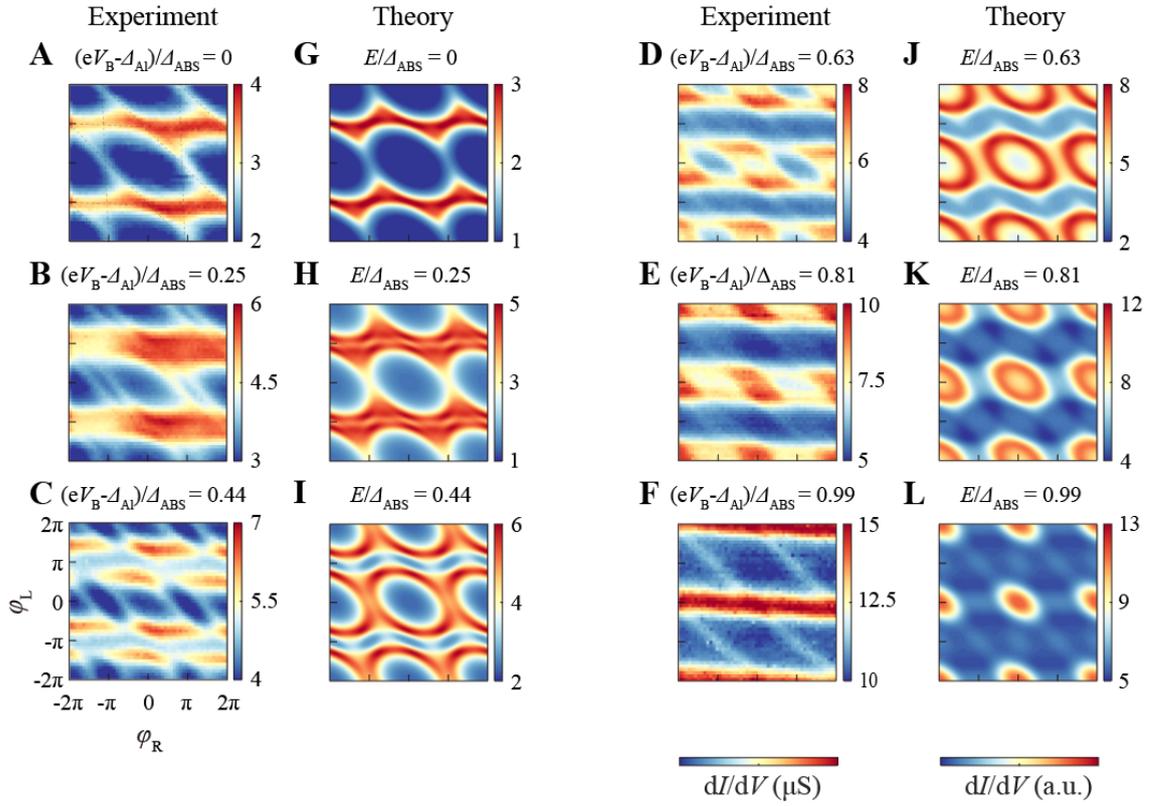

**Fig. S8. Anisotropy of R Terminal Connectivity for device 2** (**A-F**) Experimentally measured differential tunneling conductance, $dI/dV$, as a function of $(\varphi_L, \varphi_R)$, for different bias voltage, $V_B$, at gate voltage $V_G = 20$ V. (**G-L**) Theoretically calculated $dI/dV$ considering the anisotropic connectivity of R terminal, plotted as a function of $(\varphi_L, \varphi_R)$ for different normalised energy $E/\Delta_{ABS}$.

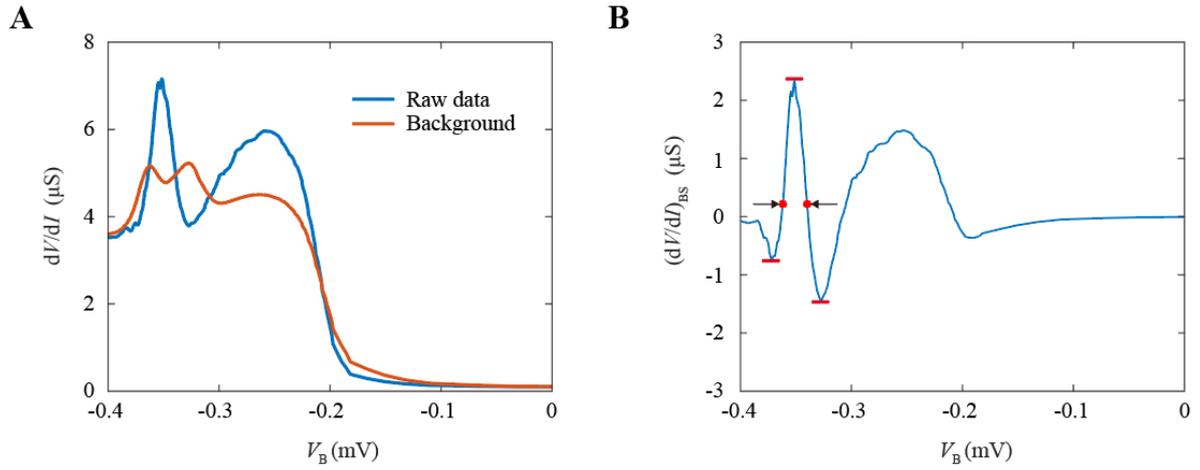

**Fig. S9. Estimated resolution of superconducting tunnel probe.** (**A**) d$I$/d$V$ versus $V_B$ for device 1 is displayed, with the raw experimental data shown in blue. The background, derived by moving-average smoothing of the d$I$/d$V$ curve, is depicted in red. (**B**) Background-subtracted differential conductance (d$I$/d$V$)$_{BS}$ relative to $V_B$ is shown. This was calculated by deducting the red line (background) from the blue line (raw data) in panel (**A**). A red bar highlights the peak or dip in the (d$I$/d$V$)$_{BS}$ curve, and black arrows mark the full-width half-maximum, which measures 22.5 $\mu V$.

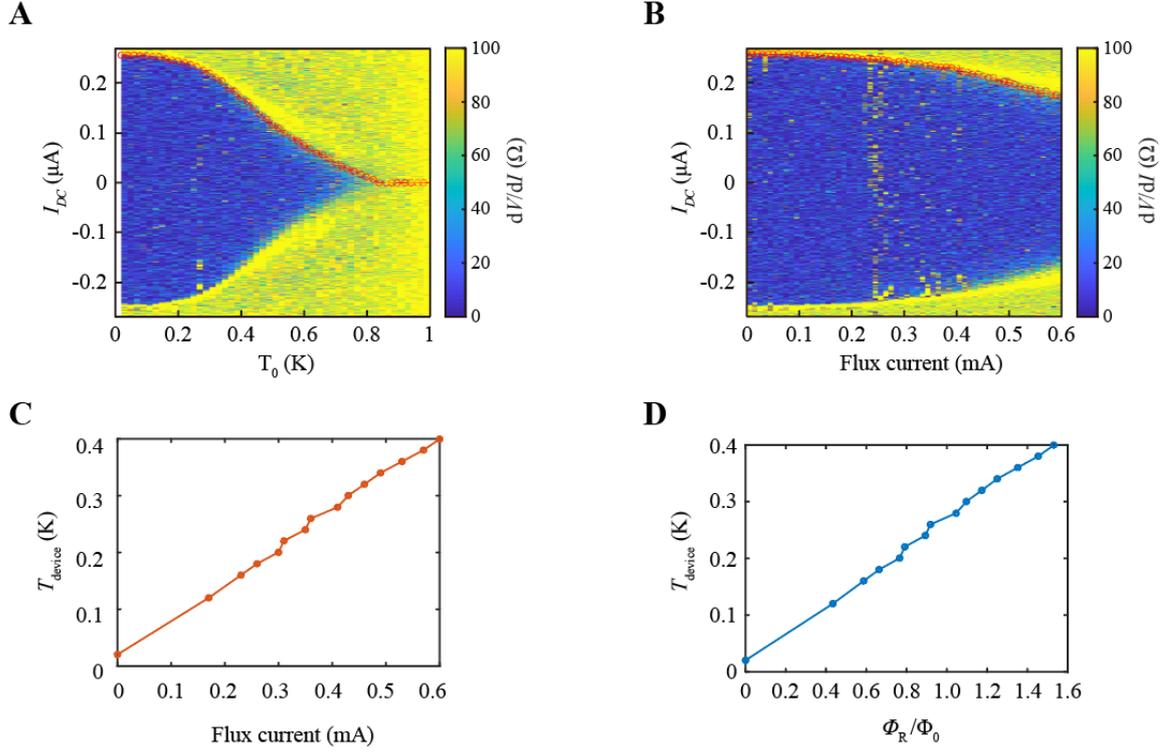

**Fig. S10. Estimation of Electron Temperature Depending on the flux gate current.**

(**A**) Differential resistance map of GJJ plotted against bias current, $I_{DC}$, and bath temperature, $T_0$, at $V_G = 20$ V. At the peak of d$V$/d$I$, highlighted with a red circle, signifies the critical current, $I_c$. As $T_0$ rises, the $I_c$ value shows a decreasing trend. (**B**) Differential resistance map of GJJ plotted against bias current, $I_{DC}$, and flux current, at $V_G = 20$ V. The $T_0$ was fixed at 20 mK. Noticeably, as the flux current grows, the $I_c$ value diminishes. (**C**) Calibrated device temperature, $T_{device}$, are presented as a function of flux current. This calibration is achieved by matching the $I_c$ values from panels (**A**) and (**B**). (**D**) Plot of the conversion from flux current to normalised magnetic flux using the periodicity of the Andreev band within the 2D phase space.